
\magnification=1200
\baselineskip=13pt
\overfullrule=0pt
\tolerance=100000

{\hfill \hbox{\vbox{\settabs 1\columns
\+ UR-1419 \cr
\+ ER-40685-868\cr
\+ hep-th/9504086\cr
}}}

\bigskip

\baselineskip=18pt

\centerline{\bf THERMAL EFFECTS ON THE CATALYSIS BY A MAGNETIC FIELD}

\bigskip
\bigskip

\centerline{Ashok Das}
\centerline{and}
\centerline{Marcelo Hott$^\dagger$}
\centerline{Department of Physics and Astronomy}
\centerline{University of Rochester}
\centerline{Rochester, NY 14627}

\bigskip
\bigskip
\bigskip
\bigskip

\centerline{\bf Abstract}

\medskip
We show that the formation of condensates in the presence of a constant
magnetic field in 2+1 dimensions is extremely unstable. It disappears as soon
as a heat bath is introduced with or without a chemical potential. We point out
some new nonanalytic behavior that develops in this system at finite
temperature.

\vskip 2 truein

\noindent $^\dagger$On leave
 of absence from UNESP - Campus de
Guaratinguet\'a, P.O. Box 205, CEP : 12.500, Guaratinguet\'a, S.P., Brazil

\vfill\eject
\noindent{\bf 1. Introduction:}

Induced quantum numbers in 2+1 dimensional quantum field theories have been
investigated in detail in the past [1-4]. There has been renewed interest in
the
subject [5-6] after the suggestion that in the presence of a constant
magnetic field,
in 2+1 dimensions, flavor symmetry is broken and fermions can generate a mass
even at the weakest attractive
interaction between the fermions [5]. More precisely, it was shown in ref. [5]
that
in the presence of a constant magnetic field, Dirac fermions (with two flavors
combined into a four component spinor) develop a nonzero value for the
condensate leading to a breakdown of the flavor symmetry. This, however,
does not give the fermions a mass. On the other hand, if one now introduces an
interaction of the Nambu-Jona Lasinio type, a fermionic mass is shown to be
generated for any value of the interaction which can be physically thought of
as
arising due to the nonzero value of the condensate in the presence of the
magnetic field.

The calculation of the condensates involves regularizing the infrared modes
with a fermion mass term which is taken to zero at the end.
In 2+1 dimensions, the value of the condensate is a discontinuous function of
the fermion mass parameter and
depends on how the zero mass limit is taken. (More explicitly,
the condensate depends on the sign of the mass term in the Lagrangian.) A
nonzero value of the condensate implies the breakdown of the flavor symmetry
(chiral symmetry) of the system.
In this short note, we will show that this formation of the condensates is a
very unstable phenomenon. The condensates disappear as soon as a heat bath is
introduced (for any finite temperature) with or without a chemical potential.
Furthermore, the theory develops a new nonanalytic behavior at the zero
temperature limit. The paper is organized as follows. In section 2, we
recapitulate briefly the calculation of the value of the condensate at
zero temperature in a manner that
extends to a finite temperature analysis naturally. In section 3, we show that
the condensate disappears for any finite value of the temperature. We point
out the
nonanalytic structure of the condensates in the limit of zero temperature. We
discuss the behavior of the thermal Bogoliubov transformations and show that
the generator of these transformations becomes nonanalytic in this limit as
well.
We present a short discussion of this behavior of the
condensates in section 4. We recognize [7] that there may be possible questions
related to the
regularization dependence of the condensates. We do not address this question
at
all. Rather, all our calculations are carried out with the zeta function
regularization.

\noindent{\bf 2. Zero Temperature:}

The standard  calculation of the condensate, in the present theory, is
carried out in a proper time
representation [5] which is not very suitable for extension to
finite temperature.
In this section, therefore, we give an alternate derivation of the zero
temperature result which readily extends to finite temperature. We follow the
notation of ref. [5] for simplicity and take the Lagrangian for the theory in
2+1
dimensions to be
$$
{\cal L} = \bar\Psi(\gamma^\mu(i\partial_\mu-eA_\mu)-m)\Psi\eqno(1)
$$
where $\Psi$ is a four component spinor involving two flavors,
$e$ is the electric charge and we choose the electromagnetic potential to
have the form $A_\mu =(0,0,Bx)$ with $B$ representing the constant, external
magnetic
field. The Landau levels can be easily calculated [8]
and have the energy values (In our entire discussions, we
will assume that $eB>0$.)
$$
E_n = \pm\sqrt{2eBn+m^2}\qquad\qquad n=0,1,2,\cdots\eqno(2)
$$
with a double degeneracy for the $n\neq 0$ modes. The degeneracy in the
$y$-component of the momentum, $p_y=p$ is also understood. The orthonormal
positive and negative energy eigenstates
can also be worked out in a straightforward manner and have the forms,
$$
\eqalign{\psi_1^{(+)}(n,p,\vec{x},t) & =
N_n\exp{(-i(|E_n|t-py))}\pmatrix{(|E_n|+m)I(n,p,x)\cr
                                \noalign{\vskip 5pt} %
                              -\sqrt{2eBn}I(n-1,p,x)\cr
                               \noalign{\vskip 5pt} %
                                 0\cr
                              \noalign{\vskip 5pt} %
                              0\cr}\cr
                               \noalign{\vskip 10pt} %
\psi_2^{(+)}(n,p,\vec{x},t) & = N_n\exp{(-i(|E_n|t-py))}\pmatrix{0\cr
                                                     \noalign{\vskip 5pt}%
                                                                 0\cr
                                                     \noalign{\vskip 5pt}%
                                                    -\sqrt{2eBn}I(n,p,x)\cr
                                                     \noalign{\vskip 5pt}%
                                                    (|E_n|+m)I(n-1,p,x)\cr}}
$$
$$\eqalign{
\psi_1^{(-)}(n,p,\vec{x},t) & =
N_n\exp{(i(|E_n|t-py))}\pmatrix{\sqrt{2eBn}I(n,-p,x)\cr
                       \noalign{\vskip 5pt}%
                              (|E_n|+m)I(n-1,-p,x)\cr
                      \noalign{\vskip 5pt}%
                               0\cr
                      \noalign{\vskip 5pt}%
                               0\cr}\cr
                       \noalign{\vskip 10pt}%
\psi_2^{(-)}(n,p,\vec{x},t) & = N_n\exp{(i(|E_n|t-py))}\pmatrix{0\cr
                                                     \noalign{\vskip 5pt}%
                                                                0\cr
                                                     \noalign{\vskip 5pt}%
                                                    (|E_n|+m)I(n,-p,x)\cr
                                                     \noalign{\vskip 5pt}%
                                            \sqrt{2eBn}I(n-1,-p,x)\cr}
}\eqno(3)$$
where
$$\eqalign{ N_n & = {1\over\sqrt{2|E_n|(|E_n|+m)}}\cr
           I(n,p,x) & = \left(eB\over\pi\right)^{1/4}{1\over\sqrt{2^n
n!}}\exp{(-{eB\over 2}(x-p/eB)^2)}H_n(\sqrt{eB}(x-p/eB))\cr
I(n=-1,p,x) & = 0
}\eqno(4)$$
with the $H_n$'s representing the Hermite polynomials. There are only two
ground state wavefunctions as can be seen from Eq. (3). Their structure depends
on the sign of the mass term. For example, for $m>0$, the two ground state
wavefunctions have the simpler form

$$\eqalign{\psi^{(+)}(0,p,\vec{x},t) & =
\exp{(-i(|E_0|t-py))}\pmatrix{I(0,p,x)\cr
                            0\cr 0\cr 0\cr}\cr
\psi^{(-)}(0,p,\vec{x},t) & = \exp{(i(|E_0|t-py))}\pmatrix{0\cr 0\cr
                                                  I(0,-p,x)\cr 0\cr}
}\eqno(5)$$

To calculate the condensate, we expand the field operator in the basis of these
wavefunctions as

$$\Psi(\vec{x},t) = \sum_n{\sum_{i=1,2}}'\int
{dp\over\sqrt{2\pi}}\;(a_i(n,p)\psi_i^{(+)}+ b_i^{\dagger}(n,p)\psi_i^{(-)})
\eqno(6)$$
where the prime in the sum over $i$ represents the fact that this sum is only
for the $n\neq 0$ modes. The creation and the annihilation operators for the
particle
and the antiparticles satisfy the standard anticommutation relations, e.g.,

$$[a_i(n,p),a_j^\dagger(n',p')]_+ =\delta_{ij}\delta_{nn'}\delta(p-p')
=[b_i(n,p),b_j^\dagger(n',p')]_+\eqno(7)$$
with all others vanishing. It is now straightforward to evaluate the condensate
which has the value

$$\langle 0|\bar\Psi(\vec{x},t)\Psi(\vec{x},t)|0\rangle = -{m\over |m|}{eB\over
2\pi}-{meB\over\pi}\sum_{n=1}^\infty{1\over E_n}\eqno(8)$$
where $E_n$ stands for the positive root given in Eq. (2). The second sum on
the right hand side is a
Hurwitz zeta function which in the limit of vanishing $m$ reduces to the
Riemann zeta function, $\zeta_R(1/2)$ [9]. The Riemann zeta function,
$\zeta_R(s)$,
is an analytic function [10] in the entire complex plane except for a simple
pole
at $s=1$. Its value
at $s=1/2$ is known to be $\zeta_R(1/2)=-1.46$ [10-11]. Thus, we see that in
the limit
of vanishing mass, the condensate has the value

$$\langle 0|\bar\Psi(\vec{x},t)\Psi(\vec{x},t)|0\rangle = -{m\over |m|}{eB\over
2\pi}\eqno(9)$$
This is the same value as in ref. [5] (except for the missing sign of the mass
factor).

\noindent{\bf 3.Finite Temperature:}

The transition to finite temperature is now straightforward. We use thermo
field
dynamics [12] for our discussion. Introducing a tilde field, we note that we
can
write a thermal doublet of fermionic fields as [12-14]

$$\Phi(\vec{x},t) = \pmatrix{\Psi(\vec{x},t)\cr \tilde\Psi^\dagger(\vec{x},t)
\cr}\eqno(10)$$
which can be expanded in terms of the wavefunctions derived earlier as

$$\Phi(\vec{x},t) = \sum_n{\sum_{1=1,2}}'\int{dp\over\sqrt{2\pi}}\left[
\pmatrix{a_i(n,p)\cr \tilde a_i^\dagger(n,p)\cr}\psi_i^{(+)}(n,p,\vec{x},t)
+\pmatrix{b_i^\dagger(n,p)\cr \tilde
b_i(n,p)\cr}\psi_i^{(-)}(n,p,\vec{x},t)\right]
\eqno(11)$$

We can now introduce the thermal Bogoliubov transformation of the form

$$U(\theta) = \exp{(iG(\theta))}\eqno(12)$$
where  the generator of the transformation is given by

$$G(\theta)=i\sum_n{\sum_{i=1,2}}'\int dp\;[\theta_i^{(+)}(n)(\tilde a_i a_i-
a_i^\dagger\tilde a_i^\dagger)+\theta_i^{(-)}(n)(\tilde b_i b_i - b_i^\dagger
\tilde b_i^\dagger)]\eqno(13)$$
and

$$\eqalign{\sin^2\theta_i^{(+)}(n) & = n_F(E_n)\cr
\sin^2\theta_i^{(-)}(n) & = 1-n_F(-E_n)\cr
n_F(E_n) & = {1\over \exp{(\beta(E_n-\mu))}+1}
}\eqno(14)$$
Here $\beta = 1/kT$ and $\mu$ represents the chemical potential.
The thermal vacuum can now be defined as

$$|0,\beta\rangle = U(\theta)|0,\tilde 0\rangle\eqno(15)$$
The thermal creation and annihilation operators, similarly, can be obtained
through the Bogoliubov transformation to be

$$\pmatrix{a_i^\beta(n,p)\cr \tilde a_i^{\beta\dagger}(n,p)\cr} =
\pmatrix{\cos\theta_i^{(+)}(n) & -\sin\theta_i^{(+)}(n)\cr
\sin\theta_i^{(+)}(n) & \cos\theta_i^{(+)}(n)\cr}
\pmatrix{a_i(n,p)\cr \tilde a_i^\dagger(n,p)\cr}\eqno(16)$$
and similarly for $b_i$ and $\tilde b_i$.

Given these, we can now calculate the value of the condensate in the thermal
vacuum simply as follows.

$$\eqalign{\langle 0,\beta|\bar\Psi\Psi|0,\beta\rangle & =\sum_n{\sum_{i=1,2}}'
\int{dp\over 2\pi}\;[\sin^2\theta_i^{(+)}(n)\bar\psi_i^{(+)}\psi_i^{(+)}+
\cos^2\theta_i^{(-)}(n)\bar\psi_i^{(-)}\psi_i^{(-)}]\cr
& = -{m\over |m|}{eB\over 2\pi}(1-{1\over\exp{(\beta(|m|-\mu))}+1}-{1\over
\exp{(\beta(|m|+\mu))}+1})\cr
& \quad-{meB\over\pi}\sum_{n=1}^\infty{1\over E_n}(1-{1\over \exp{(\beta
(E_n-\mu))}+1}-{1\over \exp{(\beta(E_n+\mu))}+1})}\eqno(17)$$
It is clear now that in the limit $m\rightarrow 0$ (for finite $\beta)$,

$$ \langle 0,\beta|\bar\Psi\Psi|0,\beta\rangle = 0\eqno(18)$$
That is, the condensate vanishes for any finite, nonzero temperature.

It is also clear now that the structure of the condensate  exhibits a
nonanalyticity at finite temperature. This is best seen by setting $\mu=0$. In
this case, we note that we can take the limit
$$
m\rightarrow 0\quad \beta\rightarrow\infty\quad |m|\beta=\alpha\eqno(19)
$$
Then, the value of the condensate, in this limit, can be obtained from
Eq. (17) to be
$$
\langle 0,\beta|\bar\Psi\Psi|0,\beta\rangle\rightarrow -{m\over |m|}{eB\over
2\pi}(1-{2\over\exp{(\alpha)}+1})\eqno(20)
$$
This shows that the order of the limits $m\rightarrow 0$ and
$\beta\rightarrow\infty$ are not commutative. That is, the condensate is not
analytic at $T=0$. Finite temperature field theories are known to exhibit
nonanalyticity [15-16], but this is a new kind of nonanalyticity, namely, the
value of the condensate is nonanalytic at the origin in the $(m, T)$ plane.

To understand the peculiar behavior of the condensate as well as the
nonanalyticity at finite temperature, we analyze next the structure of the
Bogoliubov transformation in Eqs. (12-13). And to simplify the analysis, we set
the
chemical potential $\mu=0$. The first simplification that occurs in this case
is that
$$
\theta_i^{(+)}(n) = \theta_i^{(-)}(n)
$$
and from Eq. (14), we note that there is really one $\theta$ for every mode
satisfying
$$
\tan\theta(n) = \exp{(-{\beta E_n\over 2})}\eqno(21)
$$
The generator of the Bogoliubov transformation, in this case, has the form
$$
G(\theta)=i\sum_n{\sum_{i=1,2}}'\int dp \;\theta(n)(\tilde a_i
a_i-a_i^\dagger\tilde a_i^\dagger+\tilde b_i b_i-b_i^\dagger\tilde
b_i^\dagger)\eqno(22)
$$
For $m\neq 0$, this indeed has the right behavior in that, as $\beta\rightarrow
\infty $,
$$
\tan\theta(n) =0\quad\Rightarrow \theta(n)=0\eqno(23)
$$
so that the Bogoliubov transformation simply reduces to the identity operator.
However, for any finite $\beta$ or temperature, if we let $m\rightarrow 0$, we
note from Eq. (21) as well as from the definition of $E_n$ that
$$
\tan\theta(n=0)=1\quad\Rightarrow \theta(n=0) = {\pi\over 4}\eqno(24)
$$
It is, of course, the ground state that contributes to the condensate in the
vanishing mass limit and we see that for this mode $\theta$ has the unique
value of $\pi/4$. The thermal states generated by the Bogoliubov
transformations can be thought of as squeezed states [17] which in some sense
are
polarized states. More specifically, for a four mode system, as is the case
here $(a, \tilde a, b,
\tilde b)$, it can be easily checked that in a squeezed state parameterized by
an angle $\theta$
$$
\langle\theta|a^\dagger a-bb^\dagger|\theta\rangle = -\cos 2\theta\eqno(25)
$$
Consequently, for $\theta = \pi/4$, the squeezed state acts as a crossed
polarizer for the expectation value of this operator which vanishes. (This is
the operator expectation value for the lowest mode which is responsible for the
value of the condensate and this is how the value of the condensate vanishes.)

Let us also note that in the limit of Eq. (19), all the $\theta(n)$'s vanish
except for the lowest energy mode which gives
$$
\tan\theta(n=0) = \exp{(-{\alpha\over 2})}\quad\Rightarrow
\theta(n=0)=\theta(\alpha)\eqno(26)
$$
That is, in this case, the Bogoliubov transformation does not reduce to the
identity operator and, in fact, depends on how the zero temperature and the
zero mass limit is taken. This is the nonanalyticity in the condensates that we
discussed earlier and we see that in this case, the generator of the
Bogoliubov transformation, itself, becomes nonanalytic. Consequently, not only
is the value of the condensate nonanalytic at the origin in the $(m, T)$ plane,
but most observables are likely to be. This, as we have
pointed out, is a new kind of nonanalyticity at finite temperature.

\noindent{\bf 4.Conclusion:}

In this paper, we have pointed out that the formation of condensates in a 2+1
dimensional field theory, in the presence of a constant, external magnetic
field, is a
highly unstable phenomenon. As soon as a heat bath is introduced, the value of
the condensate vanishes for any finite temperature. The thermal
expectaion value appears to average over the two possible ways of taking
the zero mass limit, namely, $m\rightarrow 0^+$ and $m\rightarrow 0^-$,
leading to a vanishing value for the condensate. Furthermore, we have
pointed out how a new kind of finite temperature nonanalyticity develops in
these theories. Namely, most observables including the value of the condensate
appear to be nonanalytic at the origin in the $(m, T)$ plane.

One of us (A.D.) would like to thank Prof. N. P. Chang for discussions.
This work was supported in part by the U.S. Department of Energy Grant No.
DE-FG-02-91ER40685. M.H. would like to thank the
 Funda\c c\~ao de Amparo a Pesquisa
do Estado de S\~ao Paulo for financial support.

\vfill\eject

\noindent {\bf References:}

\medskip

\item{1.} I. Affleck, J. Harvey and E. Witten, Nuc. Phys. {\bf B206}(1982)413.
\item{2.} A. Redlich, Phys. Rev. Lett. {\bf 52}(1984)18; Phys. Rev. {\bf
          D29}(1984)2366.
\item{3.} M. B. Paranjape, Phys. Rev. Lett. {\bf 55}(1985)2390.
\item{4.} D. Boyanovsky, R. Blanckenbecler and R. Yahalom, Nuc. Phys. {\bf
          B270}(1986)483; Phys. Rev. {\bf D34}(1986)612.
\item{5.} V. P. Gusynin, V. A. Miransky and I. A. Shovkovy, Phys. Rev. Lett.
          {\bf 73}(1994)3499.
\item{6.} D. Cangemi, E. D'Hoker and G. V. Dunne, Phys. Rev. {\bf
          D51}(1995)R2513; R. Parwani, preprint IP/BBSR/95-12, hep-th/9504020.
\item{7.} C. R. Hagen, \lq\lq Comment on \lq Catalysis of Dynamical Flavor
          Symmetry Breaking By a Magnetic...'", Univ. Rochester preprint.
\item{8.} M. Kobayashi and M. Sakamoto, Prog. Theo. Phys. {\bf 70}(1983)1375.
\item{9.} See for example, S. Blau, M. Visser and A. Wipf, Int. J. Mod. Phys.
          {\bf A6}(1991)5409.
\item{10.} H. M. Edwards, \lq\lq Riemann's Zeta Function", Academic Press, 1974
          (New York).
\item{11.} E. Jahnke and F. Emde, \lq\lq Tables of Functions", Dover
           Publications, 1956 (New York).
\item{12.} H. Umezawa, H. Matsumoto and M. Tachiki, \lq\lq Thermo Field
           Dynamics", North Holland Publishing, 1982 (Amsterdam).
\item{13.} I. Ojima, Ann. Phys. {\bf 137}(1981)1.
\item{14.} P. Elmfors, D. Persson and B. S. Skagerstam, Phys. Rev. Lett. {\bf
           71}(1993)480; see also preprint hep-ph/9312226.
\item{15.} H. A. Weldon, Phys. Rev. {\bf D26}(1982)1394; {\it ibid} {\bf
           D28}(1983)2007.
\item{16.} H. A. Weldon, Phys. Rev. {\bf D47}(1993)594; P. F. Bedaque and A.
           Das, Phys. Rev. {\bf D47}(1993)601.
\item{17.} See, for example, H. Umezawa, \lq\lq Advanced Field Theory: Micro,
Macro and Thermal Physics", American Institute of Physics, 1993 (New York).
\end